\documentclass[aps,prd,notitlepage,nofootinbib,superscriptaddress,twocolumn]{revtex4-1}

\usepackage[utf8]{inputenc}
\usepackage{amsmath}
\usepackage{amssymb}
\usepackage{amsfonts}
\usepackage{newtxtext,newtxmath}
\usepackage{bm}
\usepackage{graphicx}
\usepackage[usenames,dvipsnames]{xcolor}
\usepackage{color}
\usepackage[colorlinks=true,linkcolor=blue,urlcolor=blue,citecolor=blue]{hyperref}

\usepackage{slashed}
\usepackage[english]{babel}
\usepackage{dcolumn}
\usepackage{blindtext}
\usepackage{epsfig}
\usepackage{pifont}
\usepackage{dsfont,mathrsfs}
\usepackage{cancel}
\usepackage{bigints}
\usepackage{accents}
\usepackage{soul}
\usepackage{multirow}
\usepackage{soul}
\usepackage{simpler-wick}
\usepackage{float}

\newcommand{\sign}{\text{sign}}

\def\beq{\begin{equation}}
\def\eeq{\end{equation}}
\def\beqa{\begin{eqnarray}}
\def\eeqa{\end{eqnarray}}
\def\ban{\begin{eqnarray*}}
\def\ean{\end{eqnarray*}}
\def\bi{\begin{itemize}}
\def\ei{\end{itemize}}

\begin{document}

\title{Effects of the quark anomalous magnetic moment in the chiral symmetry restoration: magnetic catalysis and inverse magnetic catalysis}

\author{R. L. S. Farias} \email{ricardo.farias@ufsm.br}
\affiliation{Departamento de F\'{\i}sica, Universidade Federal de Santa Maria, 
97105-900 Santa Maria, RS, Brazil} 
    
\author{William R. Tavares} \email{williamr.tavares@posgrad.ufsc.br}
\affiliation{Departamento de F\'{\i}sica, Universidade Federal de Santa
  Catarina, 88040-900 Florian\'{o}polis, Santa Catarina, Brazil}  
  
\author{Rodrigo M. Nunes} \email{rodrigo.nunes@acad.ufsm.br}
\affiliation{Departamento de F\'{\i}sica, Universidade Federal de Santa Maria, 
97105-900 Santa Maria, RS, Brazil}    
 
\author{Sidney S. Avancini} \email{sidney.avancini@ufsc.br}
\affiliation{Departamento de F\'{\i}sica, Universidade Federal de Santa
  Catarina, 88040-900 Florian\'{o}polis, Santa Catarina, Brazil}

\begin{abstract}
In this work, we consider the effect of a constant anomalous magnetic moment (AMM) of quarks in the SU(2) Nambu--Jona-Lasinio 
model in the mean field approximation. To this end, we use the Schwinger {\it ansatz}, which represents a linear magnetic field term 
in the Lagrangian. A regularization method inspired in the vacuum magnetic regularization (VMR) is adopted to avoid ultraviolet 
divergences. Our results indicate a smooth decrease of the pseudocritical temperature and quark condensates for
magnetic fields $B \leq 0.1$ GeV$^2$ when a sizable AMM is considered. We found only a small window for Inverse Magnetic 
Catalysis (IMC), in contradiction with NJL predictions made in the literature. For a low value of AMM, we observe for all ranges 
of magnetic fields considered that the pseudocritical temperature increases with the magnetic field, indicating only Magnetic Catalysis 
(MC). In our approach, for nonvanishing quark AMM, the chiral symmetry restoration happens always as a smooth crossover and never 
turns into a first order phase transition.
\end{abstract}

\maketitle

\section{Introduction}

Strong magnetic fields can be generated by heavy ion collisions, with a magnitude of $eB\sim 10^{19}$ G, and this possibility has revealed an impressive amount of effects to explore. 
It can be useful to understand some fundamental aspects of quantum chromodynamics (QCD) that has general interesting, e.g, the Chiral Magnetic Effect \cite{Fukushima:2008xe,Kharzeev:2009pj,Kharzeev:2013ffa}, 
and its role in the violation of $\mathcal{P}$ and $\mathcal{CP}$ invariance. Experimental observables, as the azimuthal correlator $\gamma$ \cite{Zhao:2019hta}, are expected
to be affected by such magnetic fields. Besides, magnetars with $eB\sim 10^{16}$ G \cite{Duncan:1992hi} are natural laboratories in the universe to study such scenarios at high densities.

The role of strong magnetic fields is widely explored in effective models due its simplicity. 
Several quantities are then predicted, as the magnetic catalysis
effect (MC) \cite{Andersen:2012zc,Miransky:2015ava,Ayala:2021nhx} characterized by the increasing of the chiral condensate with the magnetic fields. 
This effect can alter many important quantities,
as the pressure, sound velocity, heat capacity of the system \cite{Farias:2016gmy,Ferreira:2013tba,Ferreira:2014kpa,Avancini:2020xqe} 
or even alter the pole-masses of light mesons \cite{Fayazbakhsh:2012vr,Avancini:2015ady,Avancini:2016fgq,Avancini:2018svs,Ayala:2020dxs,
Ayala:2018zat,Sheng:2020hge,Coppola:2018vkw,Dumm:2020muy,Mao:2018dqe}. 
There is also a possibility of such fields to alter significantly the measure of the elliptic flow $v_2$ as a direct result of the 
paramagnetic nature of QCD vacuum~\cite{Avancini:2020xqe,Bali:2013owa}.

The evaluations taken by lattice quantum chromodynamics (LQCD) confirms the MC effect at low temperatures. But a totally different
result is observed close to the pseudocritical temperature, where
the chiral condensate decreases non monotonically its value as the strong magnetic fields grow, i.e.,
for  $eB \gtrsim 0.2\text{ GeV}^2$. This is the well-know inverse magnetic catalysis effect (IMC) 
\cite{Bali:2012zg,Bali:2011qj,Bali:2013esa}, more details can be found in recent reviews \cite{Andersen:2021,Bandyopadhyay:2020zte}. 
Such evaluations are considered with physical pion masses \cite{Bali:2012zg}, where it is also observed the decreasing of the pseudocritical temperature.
For heavy pions, the chiral condensate suffers a MC, but the decreasing of the 
pseudocritical temperature still persists \cite{Endrodi:2019zrl} .

There are several attempts trying to explain such a contradiction between LQCD results and the lack of theoretical predictions for IMC using 
effective models of QCD. See Refs \cite{Andersen:2012zc,Miransky:2015ava} 
for reviews. Some proposals explore
the evaluations of the Nambu--Jona-Lasinio (NJL) model beyond mean field approximation \cite{Mao:2016fha}, 
others study the implementation of a magnetic or thermo-magnetic dependence of the coupling constant of the 
model fitted by LQCD data 
\cite{Farias:2014eca,Farias:2016gmy,Endrodi:2019whh,Moreira:2020wau,Moreira:2021ety,Tavares:2021fik} 
and with the implementation of the anomalous magnetic moment (AMM)
of quarks \cite{Fayazbakhsh:2014mca,Mei:2020jzn,Chaudhuri:2019lbw,Chaudhuri:2020lga}.

The study of AMM of quarks has started more than twenty years ago, and several approaches were used \cite{Bicudo:1998qb}.
But, the relation of the AMM with IMC is calling the attention very recently by some authors 
\cite{Fayazbakhsh:2014mca,Mei:2020jzn,Chaudhuri:2019lbw,Chaudhuri:2020lga,Ghosh:2020xwp}. 
It is possible to draw a QCD phase diagram with the influence of the AMM phenomenologically estimated by quark models \cite{Fayazbakhsh:2014mca} 
or, even, to evaluate
the electromagnetic vertex function to predict the value of AMM as a function dependent on the temperature and magnetic fields \cite{Ghosh:2021dlo}. 
The mesonic excitations,
as well as thermodynamic properties, have been explored with the AMM influence \cite{Chaudhuri:2019lbw,Chaudhuri:2020lga,Xu:2020yag}. 
Also, some results at finite density with light mesons or thermodynamic properties are reported by \cite{Aguirre:2020tiy,Aguirre:2021ljk}.

In the context of the NJL model \cite{Nambu:1961tp,Nambu:1961fr} and its extensions, as de PNJL \cite{Ratti:2005jh},
the four-fermion interaction channel provides a non-renormalizable theory at $3+1$ D, and one needs to adopt a regularization prescription to avoid
ultraviolet divergences.
In this way, based on well  known results of \cite{Ebert:1999ht,Ebert:2003yk,Menezes:2008qt},
and recently explored in Ref. \cite{Avancini:2019wed}, the magnetic field independent regularizations (MFIR) present to be satisfactory
when treating the implementation of a constant external magnetic field. This regularization prescription completely separates the magnetic field 
contributions from the vacuum or thermal quantities. On the other hand, most of the non-MFIR based regularizations showed 
unphysical results \cite{Allen:2015paa,Duarte:2015ppa}, as oscillations in the chiral quark condensate \cite{Avancini:2019wed}
or tachyonic neutral pion masses \cite{Fayazbakhsh:2012vr}. There are also other regularizations that can
avoid these oscillations, as the Pauli-Villars scheme \cite{Chaudhuri:2021skc,Mao:2016fha,Mao:2018dqe} or the Vacuum magnetic regularization scheme \cite{Avancini:2020xqe,Tavares:2021fik}.

To include the AMM of quarks, most of the cited works
made use of the linear magnetic field {\it ansatz} proposed by Schwinger in the context of quantum electrodynamics (QED) \cite{Schwinger:1948iu}. 
 However, in the context of finite temperatures and zero density, none of these works apply a proper separation of the pure magnetic contributions from the vacuum, like MFIR or VMR regularization 
schemes.  In a QED framework, it is remarkable the results of Ref. \cite{Dittrich:1977ee}, where the one-loop effective potential with the AMM of the electron added through the Pauli term
is investigated. Moreover, an exact expression up to the order $\alpha^2$ is achieved. The range of applicability of this approach  was discussed for the electron-positron pair-production \cite{Dittrich:1981ir,OConell:1968,Jancovici:1970ep}, as well as when applied to quarks \cite{Ferrer:2015wca}.
Other possibilities are also explored, as in Ref. \cite{Ferrer:2013noa}, where the quarks' masses and the AMM appear as a result of a dynamical 
generation associated with a scalar and an appropriated tensor channel in the Lowest Landau Level approximation. 

In this work, we consider in an effective way the effects of the AMM of the quarks in a magnetized medium through the addition of the phenomenological Pauli term to the SU(2) NJL model. In order to treat the vacuum divergent term, we make use of an ingenious formalism developed in the context of the QED theory in ref. \cite{Dittrich:1977ee} to incorporate the electron AMM effectively where the vacuum term is given as an analytic expression similar to 
the one obtained by Euler-Heisenberg \cite{Heisenberg:1936nmg}, Weisskop \cite{Weisskopf:1936hya} and Schwinger \cite{Schwinger:1951nm}, where the usual summation on the Landau levels is performed in an exact way. We adapt the latter formalism originally proposed for the renormalizable QED theory for the non-renormalizable SU(2) NJL model. The divergent effective potential is treated by using an appropriate subtraction scheme where we separate clearly and analytically the convergent term from the divergent ones. In contrast to renormalizable theories, the way the divergences are treated is to be considered as a part of the definition of the model. We choose to use the 3D cutoff technique in the VMR which has been proven to be reliable \cite{Avancini:2019wed,Avancini:2020xqe} for the NJL model. In fact, this formalism has been shown to guarantee that the Gell-Mann-Oakes-Renner relation is respected \cite{Avancini:2020xqe}. One of the main objectives of the present work is to show the importance of a proper regularization procedure. In the literature several ways have been used to deal with the infinities inherent to the non-renormalizable NJL model, such as Form Factors \cite{Fayazbakhsh:2014mca}, B-dependent cutoff \cite{Chaudhuri:2019lbw}, Pauli-Villars \cite{Chaudhuri:2021skc}. Besides the infinite vacuum contribution, often in previous works the finite medium term has also being affected by the regularization procedure, which has been shown to be an inadequate procedure \cite{Avancini:2019wed}. 

The inclusion of AMM effects depends on the choice of the anomalous magnetic moments of the quarks. We use two sets of AMM values given in the literature \cite{Fayazbakhsh:2014mca}. Here, our focus is to analyze the behavior of the effective quark mass as a function of temperature, magnetic field strength and its dependence on the AMM of the quarks. In addition, the importance of the regularization procedure will be emphasized. Our final aim is to study the behavior of the pseudo-critical temperature as a function of the magnetic field and the influence of the AMM in this case. This will allow us to see if the mechanism of the inverse magnetic catalysis can be associated to the inclusion of AMM of quarks, as has been reported in the recent literature. As we have already  mentioned before, the formalism adopted here has its limit of validity that, of course, has been taken into account in our calculations.

 This work is organized as follows. In Sec.II we present the SU(2) NJL model with the AMM term included.  In Sec.III, we develop the formalism, in Sec.IV
we obtain the GAP equation. Numerical results are presented in Sec. V and our the conclusions in Sec. VI. In the appendix A and B specific technical details are discussed.

\section{Lagrangian of the SU(2) NJL model with AMM}

The lagrangian of the SU(2) NJL model with anomalous magnetic moment in an external electromagnetic field is given by the following expression \cite{Fayazbakhsh:2014mca}

\begin{eqnarray}
\mathcal{L}&=&\overline{\psi}\left(i \slashed D - \hat{m}+\frac{1}{2}\hat{a}\sigma^{\mu\nu}F^{\mu\nu}\right)\psi\nonumber\\
           &&+G\left[(\overline{\psi}\psi)^{2}+(\overline{\psi}i\gamma_{5}\vec{\tau}\psi)^{2}\right],
\end{eqnarray}

\noindent where $A^\mu$, $F^{\mu\nu} = \partial^\mu A^\nu - \partial^\nu A^\mu$ 
are respectively the electromagnetic gauge and  tensor fields, $G$ represents the coupling 
constant, $\vec{\tau}$ are isospin Pauli matrices,  $Q$ is the diagonal quark charge 
\footnote{Our results are expressed in Gaussian natural units 
where $1\,{\rm GeV}^2= 1.44 \times 10^{19} \, G$ and $e=1/\sqrt{137}$.} matrix, 
$Q$=diag($q_u$= $2 /3$, $q_d$=-$1/3$),
$D^\mu =(\partial^{\mu} + i e Q A^{\mu})$ is the covariant derivative,   
 $\psi$ is the quark fermion field, and $\hat{m}$ represents the bare quark mass matrix,
\begin{equation}
 \psi = \left(
\begin{array}{c}
\psi_u  \\
\psi_d \\
\end{array} \right) ~, ~
 \hat{m}  = \left(
\begin{array}{cc}
m_u & 0 \\
0 & m_d \\
\end{array} \right) ~.
\end{equation}
We consider here $m_u$=$m_d=m$  and choose the Landau gauge, $A^{\mu}=\delta_{\mu 2}x_{1}B$, which satisfies 
$\nabla \cdot \vec{A}=0$ and $\nabla \times \vec{A}=\vec{B}=B{\hat{e_{3}}}$, i. e., resulting in a constant magnetic field in the z-direction. 

In the mean field approximation, the lagrangian $\mathcal{L}$ is denoted by

\begin{equation}
 \mathcal{L}=\overline{\psi}\left(i\slashed{D}-M+\frac{1}{2}\hat{a}\sigma_{\mu\nu}F^{\mu\nu}\right)\psi - \frac{(M-m)^2}{4G},
\end{equation}
\noindent where the constituent quark mass is defined by 
\begin{equation}
 M=m-2G \left \langle \overline{\psi}\psi \right \rangle. \label{gap}
\end{equation}

\noindent where $\left \langle \overline{\psi}\psi \right \rangle$ is the chiral quark condensate. The AMM factor $\hat{a}$ is given by $\hat{a}$=diag($a_u$, $a_d$) with 
$a_f=q_f\alpha_f\mu_B$, where $f={u,d}$ represents the quark flavor. 
In the one-loop level approximation, the previous quantities are given by

\begin{eqnarray}
\alpha_f=\frac{\alpha_e q_f^2}{2\pi}, \quad  \alpha_e=\frac{1}{137},\quad \mu_B=\frac{e}{2M}.
\end{eqnarray} 

\section{Effective Potential with AMM}

Using the approach adopted in Ref.~\cite{Dittrich:1977ee}, we adapt the non-regularized QED effective lagrangian in a 
constant external magnetic field with the AMM of the electron, in the one-loop approximation, to the SU(2) NJL model.  
For this purpose, we can write

\begin{eqnarray}
 \mathcal{L}(B)&=&\sum_{f=u,d}\frac{N_c}{8\pi^2}\int_0^{\infty} \frac{ds}{s^3}e^{-i\mathcal{K}_{0f}^2s}\frac{q_feBs}{\sin(q_feBs)}\nonumber\\
 &&\times \cos(2M\eta_fBs),
 \label{LAMM}
\end{eqnarray}
\noindent where we have adopted the following definitions 

\begin{eqnarray}
 \mathcal{K}_{0f}^2&=&M^2+a_f^2B^2,\label{k0f}\\\
 \eta_f&=&-q_f(\alpha_f+1)\mu_B.
\end{eqnarray}

\noindent Using the transformation $s\rightarrow -is$ in eq. (\ref{LAMM}) and defining $q_feB=B_f$, we achieve

\begin{eqnarray}
  \mathcal{L}(B)&=&\sum_{f=u,d}\frac{N_c}{8\pi^2}\int_0^{\infty} \frac{ds}{s^3}e^{-\mathcal{K}_{0f}^2s}\frac{B_fs}{\sinh(B_fs)}\nonumber\\
 &&\times\cosh[(\alpha_f+1)B_fs]. \nonumber\\\label{LAMM1}
\end{eqnarray}

The expression for $\mathcal{L}(B)$ is clearly divergent and we will apply the vacuum magnetic regularization (VMR) scheme to regularize this quantity. After some simple steps, that
are described in the Appendix \ref{appA}, we rewrite $\mathcal{L}(B)$ as 
\begin{eqnarray}
\mathcal{L}(B)=\Omega^{mag}+\Omega^{vac}+\Omega^{field}, 
\end{eqnarray}

\noindent where:

\begin{eqnarray}
\Omega^{mag}&=& \sum_{f=u,d}\frac{N_c}{8\pi^2}\int_0^{\infty} \frac{ds}{s^3}e^{-s\mathcal{K}_{0f}^2}\left\{ \frac{B_fs\cosh[(\alpha_f+1)B_fs]}{\sinh(B_fs)}\right.\nonumber\\
             &&- \left. 1-\frac{1}{6}[3(\alpha_f+1)^2-1](B_fs)^2 \right\},\label{omegamag1}
\end{eqnarray}

\noindent The contributions $\Omega^{vac}$ and $\Omega^{field}$ must be regularized. We choose the
3D sharp cutoff scheme and the following expressions are given by

\begin{eqnarray}
 \Omega^{vac}&=&-\frac{N_c}{8\pi^2}\sum_{f=u,d} \left\{ 
 \Lambda[\Lambda^2+\epsilon_f^2(\Lambda)]\epsilon_f(\Lambda)\nonumber\right.\\
&&-\left. \mathcal{K}_{0f}^4\ln\left[\frac{\Lambda+\epsilon_f(\Lambda)}{\mathcal{K}_{0f}}\right]\right\},
 \end{eqnarray}

\noindent and for $\Omega^{field}$

\begin{eqnarray}
 \Omega^{field}=-\sum_{f=u,d}\frac{N_cB_f^2}{48\pi^2}\left[3(\alpha_f+1)^2-1\right] \ln\frac{\mathcal{K}_{0f}^2}{\Lambda^2}.
\end{eqnarray}

\noindent where $\epsilon_f^2(\Lambda)=\mathcal{K}_{0f}^2+\Lambda^2$. 
We will find, at the end of the derivations, the following thermodynamic potential of the SU(2) NJL model 

\begin{eqnarray}
 \Omega &=&\frac{(M-m)^2}{4G}+\Omega^{vac}+\Omega^{field}\nonumber \\
 && + \Omega^{mag}+\Omega^{Tmag} + \frac{1}{2}B^2,
        \label{omegafull}
\end{eqnarray}

\noindent where $\Omega^{Tmag}$ is the finite 
thermo-magnetic contribution of the medium, given by the following expression:

\begin{eqnarray}
\Omega^{Tmag}&=&-\frac{1}{\beta}\sum_{f=u,d}\frac{N_c|B_f|}{4\pi^2}\sum_{n=0}^{\infty}\sum_{s=\pm 1}\int_{-\infty}^{\infty}dp_3\nonumber\\
&& \times\left[\ln\left(f^{+}(E^{f}_{n,s})\right)+\ln\left(f^{-}(E^{f}_{n,s}\right)\right],
\end{eqnarray}

We have defined the following
function $f^{\pm}(E^{f}_{n,s})=1+e^{-\beta(E^{f}_{n,s} \mp \mu_f)}$, where the inverse of the temperature is
defined as $\beta=\frac{1}{T}$ and $\mu_f$ is the chemical potential for a quark of flavor $f$.
Differently from the former quantities, the thermo-magnetic contribution is written in terms of a summation of the Landau Levels, $n$, and
the energy levels of the quarks, $E^{f}_{n,s}$, given by.

\begin{eqnarray}
 E^{f}_{n,s}=\sqrt{p_3^2+(M_{n,s}^f-sa_fB)^2},
\end{eqnarray}

\noindent with $M_{n,s}^f=\sqrt{|B_f|(2n+1-s_fs)+M^2}$, where $s_f=\sign(q_f)$ being $\sign(x)$ the signal function of $x$.

\section{The gap equation}

To evaluate the effective quark masses under external constant magnetic fields, we evaluate $ \frac{\partial \Omega}{\partial M}=0$.
The gap equation can be written as:

\begin{eqnarray}
 &&\frac{M-m}{2G}-\sum_{f=u,d}[h^{vac}_f(M)-h^{field}_f(M,eB)\nonumber\\
 &&-h^{mag}_f(M,eB)-h^{Tmag}_f(M,eB,T)]=0.\label{gap1}
\end{eqnarray}
Each piece of the above equation is defined as follows. The vacuum-magnetic $h^{vac}_f(M)$ contribution is given by

\begin{eqnarray}
 h^{vac}_f(M)= \frac{MN_c}{2\pi^2}\left\{\Lambda\epsilon_f(\Lambda)-\mathcal{K}_{0f}^2\ln\left[\frac{\Lambda+\epsilon_f(\Lambda)}{\mathcal{K}_{0f}}\right] \right\},\nonumber\\
\end{eqnarray}

\noindent The remaining magnetic field dependent contributions at $T=0$,  $h^{field}_f(M,eB)$, and $h^{mag}_f(M,eB)$ are given by

\begin{eqnarray}
h^{field}_f(M,eB)=\frac{MN_cB_f^2}{24\pi^2}\frac{[3(\alpha_f+1)^2-1]}{\mathcal{K}_{0f}^2},
\end{eqnarray}

\begin{eqnarray}
h^{mag}_f(M,eB)&=&\frac{MN_c}{4\pi^2}\int_{0}^{\infty}\frac{ds}{s^2}e^{-s\mathcal{K}_{0f}^2}\nonumber\\
               &&\times\left\{\frac{B_fs\coth[(\alpha_f+1)B_fs]}{\sinh(B_fs)}\right. \nonumber\\ 
               &&-\left.1-\frac{1}{6}\left[3(\alpha_f+1)^2-1\right](B_fs)^2\right\}.\nonumber\\ 
\end{eqnarray}

\noindent At last, the thermal $h^{Tmag}_f(M,eB,T)$ contribution is, therefore, defined as

\begin{eqnarray}
 h^{Tmag}_f(M,eB,T)&=&-\frac{MN_c|B_f|}{4\pi^2}\sum_{n=0}^{\infty}\sum_{s=\pm 1}\int_{-\infty}^{\infty}\frac{dp_3}{E_{n,s}^f}\nonumber\\
                    &&\times\left(\frac{1}{1+e^{\beta(E_{n,s}^f-\mu_f)}}+\frac{1}{1+e^{\beta(E_{n,s}^f+\mu_f)}}\right)\nonumber\\
                    &&\times\left(1-\frac{sa_fB}{M_{n,s}^f}\right).
\end{eqnarray}

Solving the eq. (\ref{gap1}) we obtain the effective quark masses with the influence of AMM. To this end, we adopt the
criteria of Ref. \cite{Fayazbakhsh:2014mca}, where the anomalous magnetic moment is considered as a constant value through the relation $k_f=\alpha_f/2M$.
Therefore, $\mathcal{K}_{0f}$ in eq. (\ref{k0f}) is
redefined as
\begin{eqnarray}
 \mathcal{K}_{0f}=M^2+k_f^2B_f^2.
\end{eqnarray}

The last equation results from the definition $a_f=q_f\alpha_f\mu_B\rightarrow a_f=q_f ek_f$. 

\section{Results}
In this work we adopt the following set of parameters:  $\Lambda=591.6$ MeV, $m=5.7233$ MeV and $G=2.404/\Lambda^2$.  These parameters are chosen such that we have the following values in the vacuum: pion mass $m_{\pi}=138$ MeV,  pion decay constant $f_\pi=92.4$ MeV, and
the chiral condensate $\langle\overline{u}u\rangle=(-241\text{MeV})^3$ \cite{Avancini:2019wed}. We adopt two sets of parameters for
the constant values of $k_f$ that are obtained in \cite{Fayazbakhsh:2014mca}. In the first set $\kappa^{[1]}$, 
we consider the sizable quark AMM values
\begin{eqnarray}
 &&k_u^{[1]}=0.29016\hspace{0.4em} \text{GeV}^{-1},\quad k_d^{[1]}=0.35986\hspace{0.4em} \text{GeV}^{-1},\nonumber\\
 &&\alpha_u^{[1]}=0.242, \quad\quad\quad\quad\quad \alpha_d^{[1]} = 0.304,\label{set1}
\end{eqnarray}

\noindent and the second set $\kappa^{[2]}$ is given by
\begin{eqnarray}
&& k_u^{[2]}=0.00995\hspace{0.4em} \text{GeV}^{-1},\quad k_d^{[2]}=0.07975\hspace{0.4em} \text{GeV}^{-1},\nonumber\\
&& \alpha_u^{[2]}=0.006, \quad\quad\quad\quad\quad \alpha_d^{[2]} = 0.056,\label{set2}
\end{eqnarray}

\noindent The details of this parametrization can be seen in Ref.~\cite{Fayazbakhsh:2014mca,Bicudo:1998qb}.

\begin{figure}[h]
\begin{tabular}{ccc}
\includegraphics[width=8.5cm]{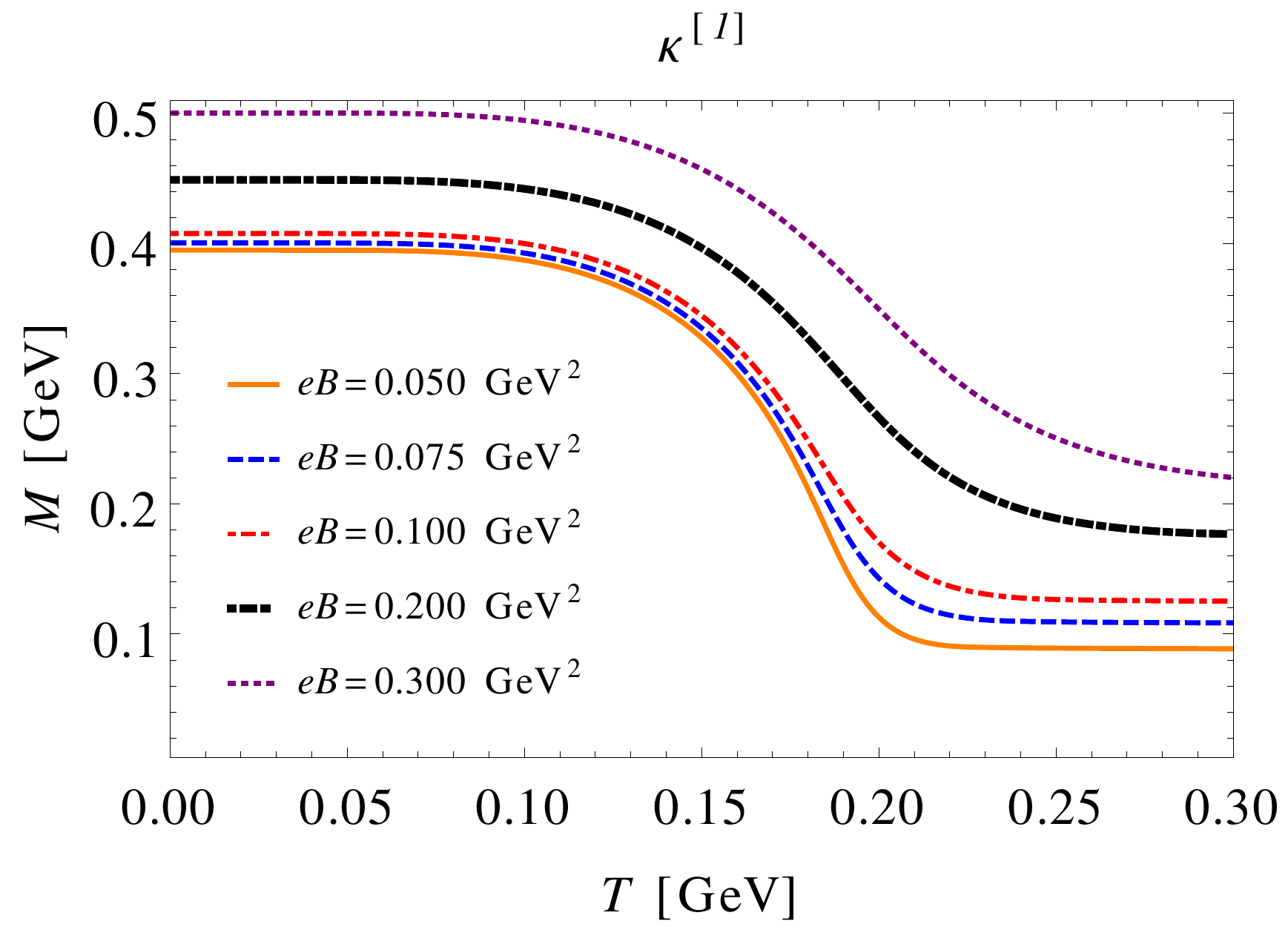}\\
\end{tabular}
\caption{Effective quark mass as a function of the temperature for different values of fixed magnetic field strength. For these results, it was adopted
the set $\kappa^{[1]}$.}
\label{MxT_kappa1}
\end{figure}

In Fig.~\ref{MxT_kappa1}, we show the effective quark mass as a function of the temperature for fixed values of the magnetic field using the set $\kappa^{[1]}$.
As observed, the partial chiral symmetry restoration is not fully obtained for all $eB$ values considered. The limit $M\approx m$ is not
reached at temperatures higher than the pseudocritical temperature, $T>>T_{c}$. This is a pure effect of the lower bound in the effective quark masses generated by the constrain obtained in Appendix \ref{appB}. For instance, for $eB=0.1$ GeV$^2$, $M\gtrsim 120$ MeV.
The lower bound on the effective quark masses 
increases as we grow the magnetic field strength. 

\begin{figure}[h]
\begin{tabular}{ccc}
\includegraphics[width=8.5cm]{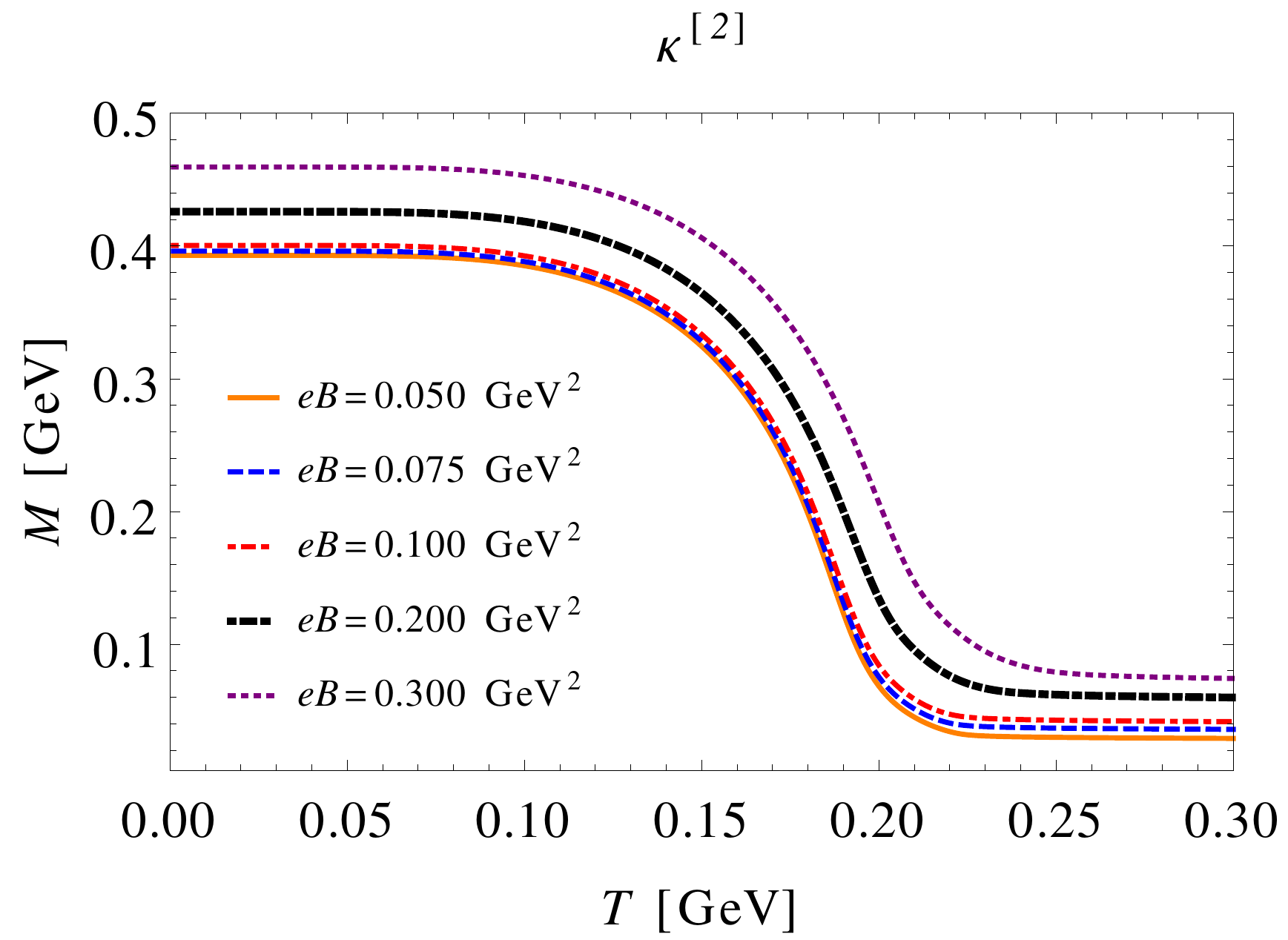}\\
\end{tabular}
\caption{Effective quark mass as function of the temperature for different values of fixed magnetic field strength. For these results, was adopted
the set $\kappa^{[2]}$.}
\label{MxT_kappa2}
\end{figure}

We show in Fig.~\ref{MxT_kappa2} as in Fig.~\ref{MxT_kappa1} the behavior of the effective quark masses as a function of the temperature for fixed $eB$, fixing in this case $\kappa=\kappa^{[2]}$. The inferior limit of the effective
quark masses are lower when compared with the results of the $\kappa^{[1]}$ set, therefore, the chiral symmetry is almost restored in the high temperature regime 
through a smooth crossover. 
It is clear from Fig.\ref{MxT_kappa1} and Fig.\ref{MxT_kappa2} that for both sets used, the magnetic catalysis effect for $T=0$, 
in contradiction with the results presented in Ref.~\cite{Chaudhuri:2019lbw,Ghosh:2020xwp}. 
Also, the type of transition that we observe is a smooth crossover, differently from the first order observed in Ref.~\cite{Fayazbakhsh:2014mca}. In Ref.~\cite{Mei:2020jzn}, the authors also observe a first order phase transition, but they consider confinement effects through the Polyakov loop in their calculations.

\begin{figure}[h]
\begin{center}
\includegraphics[width=0.48\textwidth]{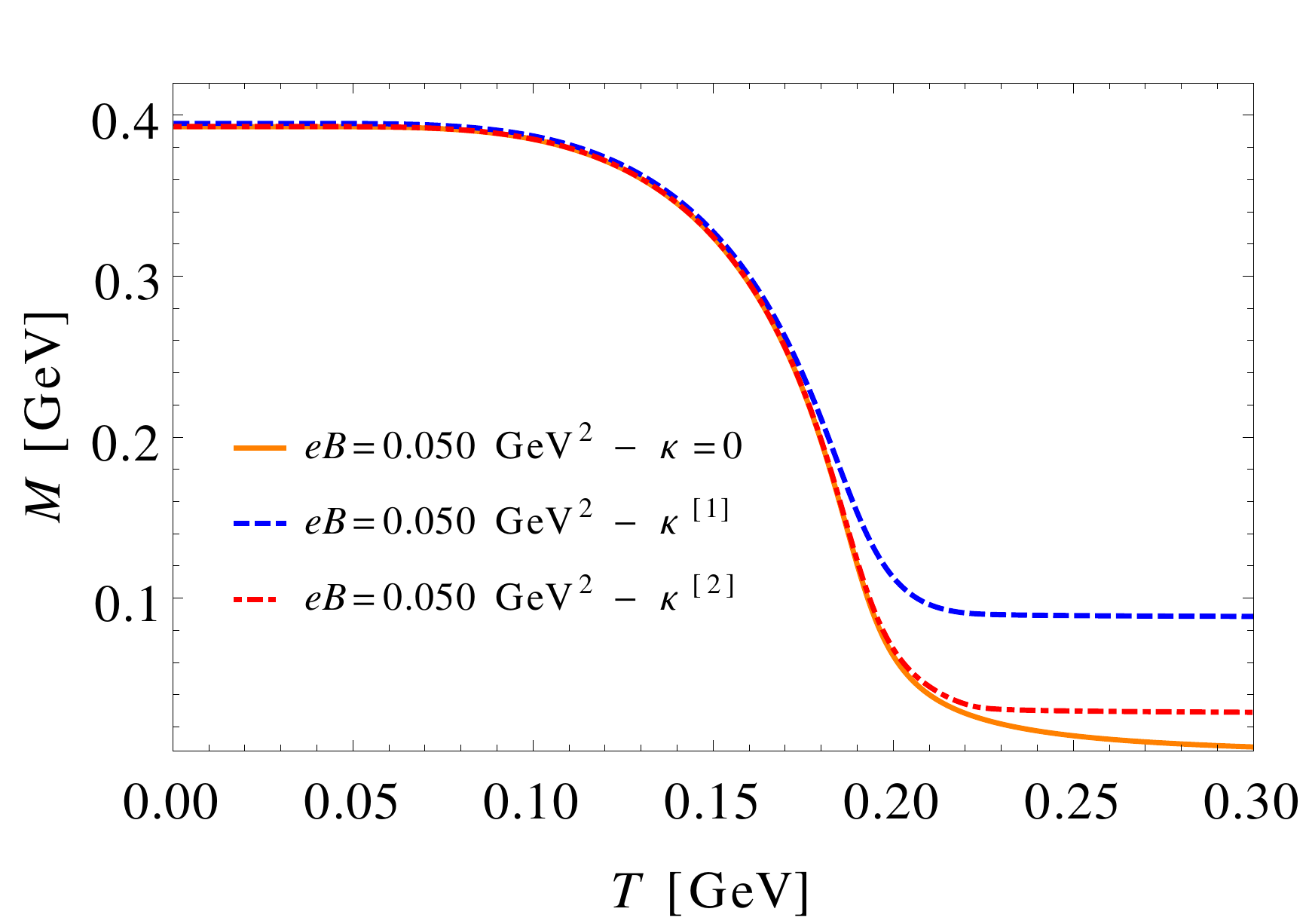}\vspace{0.5 cm}
\includegraphics[width=0.48\textwidth]{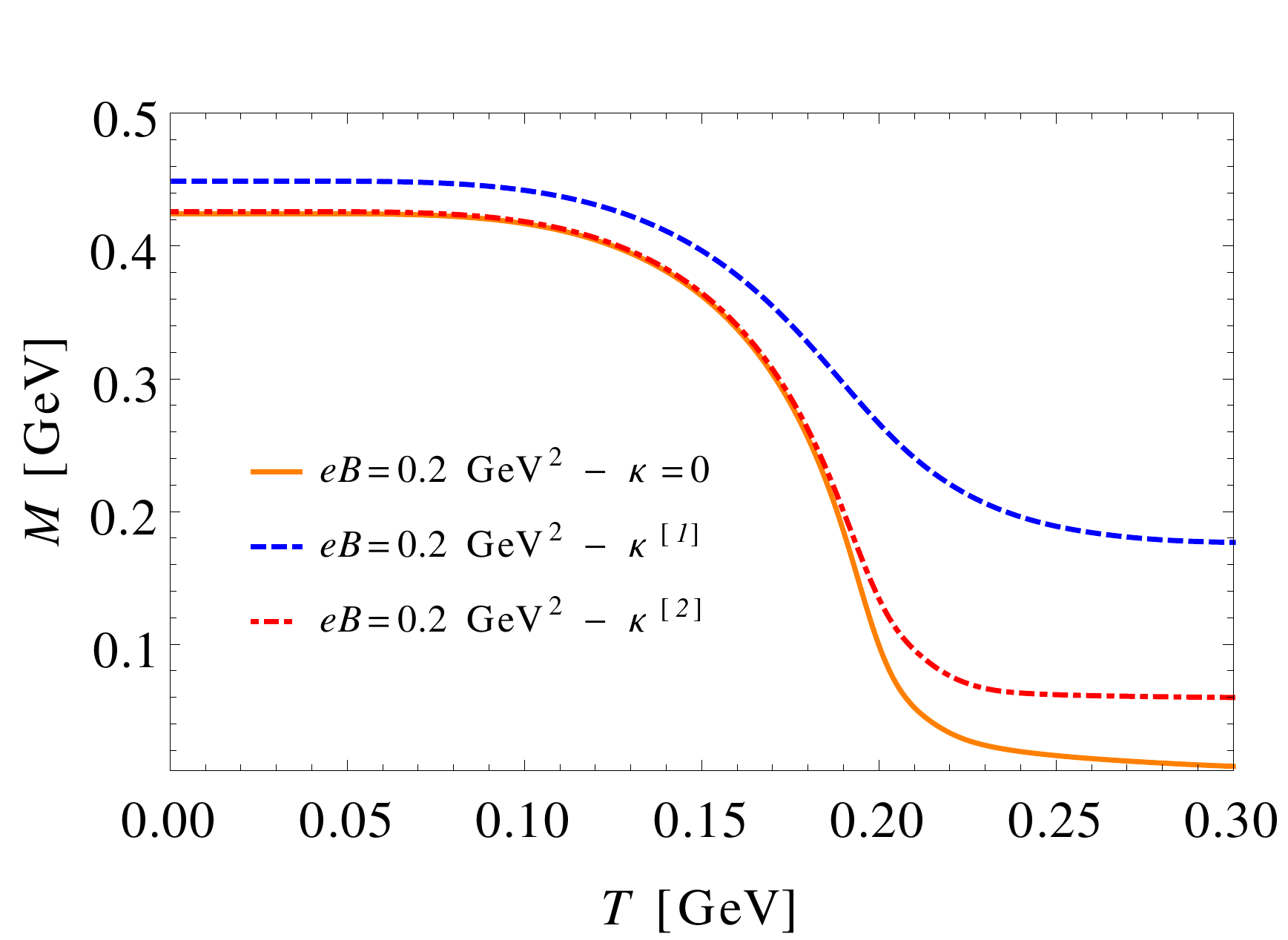}
\end{center}
\caption{Effective quark masses as a function of the temperature for $eB=0.05$ GeV$^2$ (top panel) and $eB=0.2$ GeV$^2$ (low panel) with fixed values of $\kappa$.}
\label{MxT_kappa}
\end{figure}

We present in Fig.~\ref{MxT_kappa} the effective quark mass varying with the temperature for $\kappa=0$, $\kappa=\kappa^{[1]}$ and $\kappa=\kappa^{[2]}$.
In the top panel of Fig.~\ref{MxT_kappa}, we fix $eB=0.05$ GeV$^2$. All the three curves shows similar results for $T<200$ MeV. On the other hand, the lower bound of the $\kappa^{[1]}$ and $\kappa^{[2]}$
determines the limit of the effective quark mass, and different values are obtained in the region $T>200$ MeV. In the low panel, we show the same quantities with the magnetic field,
$eB=0.2$ GeV$^2$, fixed. As we can see at low temperatures, the results show a clear difference, with $M_{\kappa^{[1]}}> M_{\kappa^{[2]}}>M_{\kappa=0}$. At high temperatures,
the lower bounds strongly determines the limit of the effective quark masses in the partial chiral symmetry restoration phase, which is not fully completed for $\kappa^{[1]}$ and 
$\kappa^{[2]}$.

\begin{figure}[h]
\begin{tabular}{ccc}
\includegraphics[width=8.0cm]{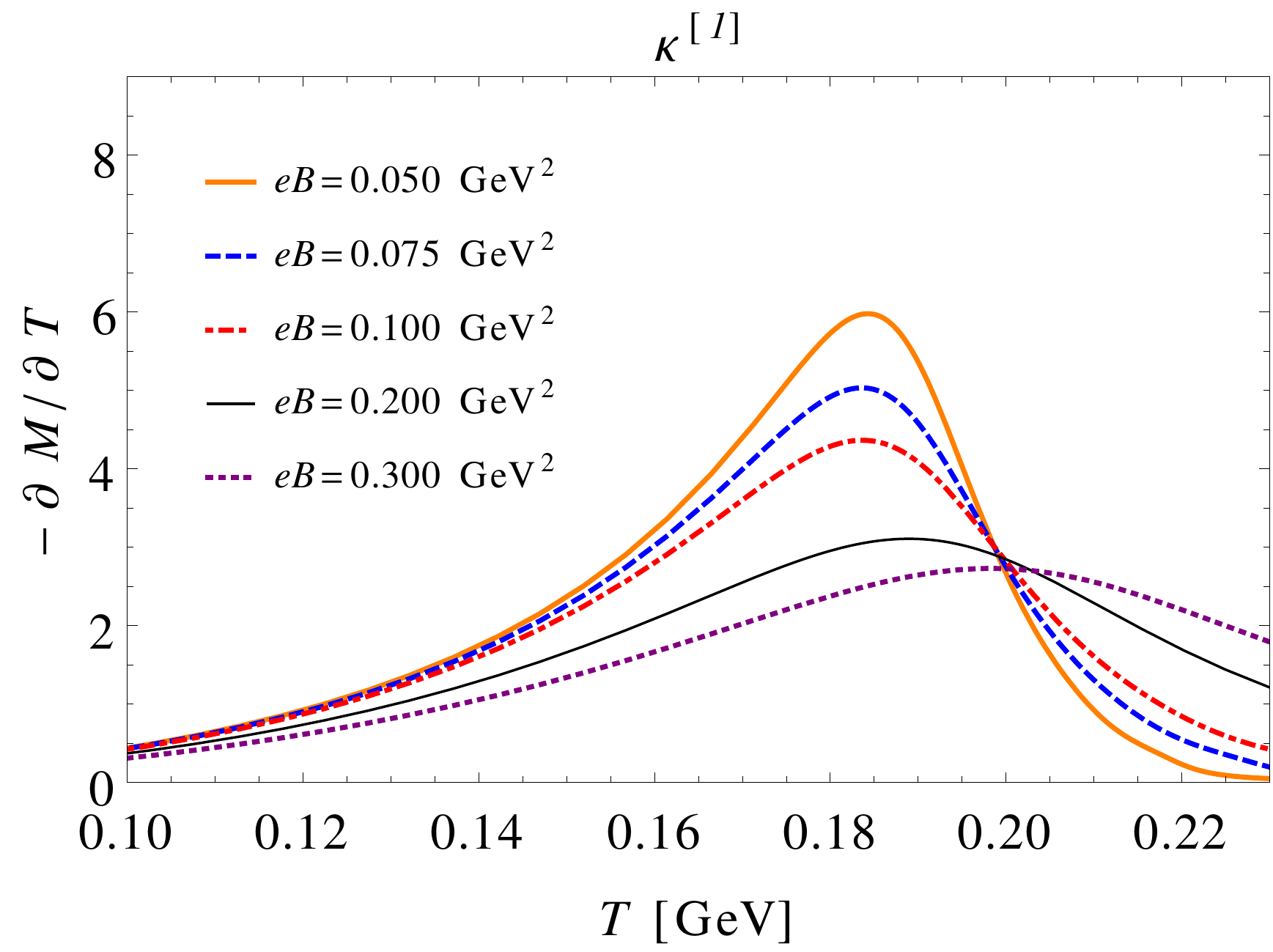}\\
\end{tabular}
\caption{$-\partial M/\partial T$ as function of the temperature, for different fixed values of magnetic field with fixed $\kappa^{[1]}$.}
\label{dmdt_kappa1}
\end{figure}

\begin{figure}[h]
\begin{tabular}{ccc}
\includegraphics[width=8.0cm]{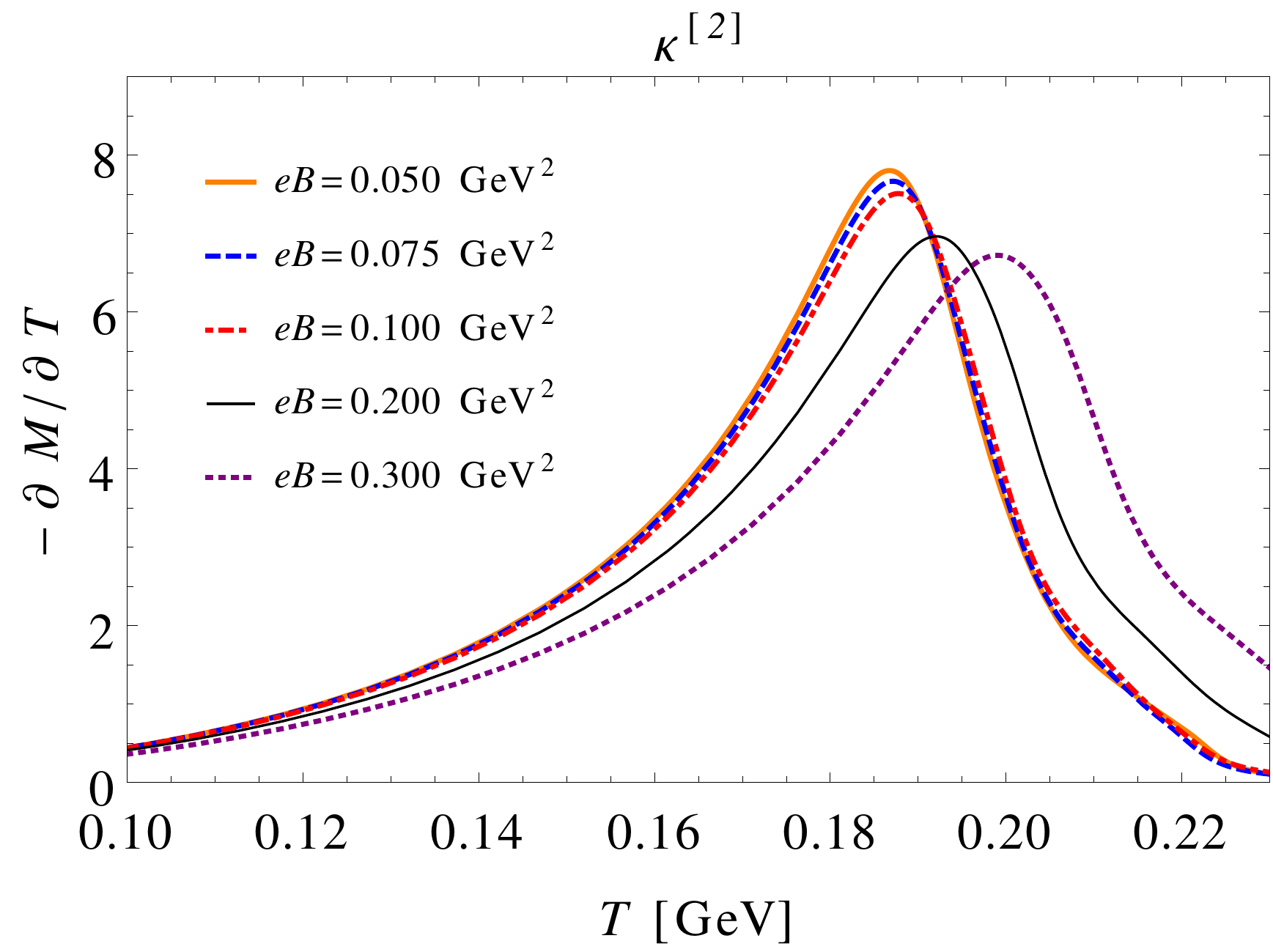}\\
\end{tabular}
\caption{$-\partial M/\partial T$ as function of the temperature, for different fixed values of magnetic field with fixed $\kappa^{[2]}$.}
\label{dmdt_kappa2}
\end{figure}
\begin{figure}[h]
\begin{tabular}{ccc}
\includegraphics[width=8.5cm]{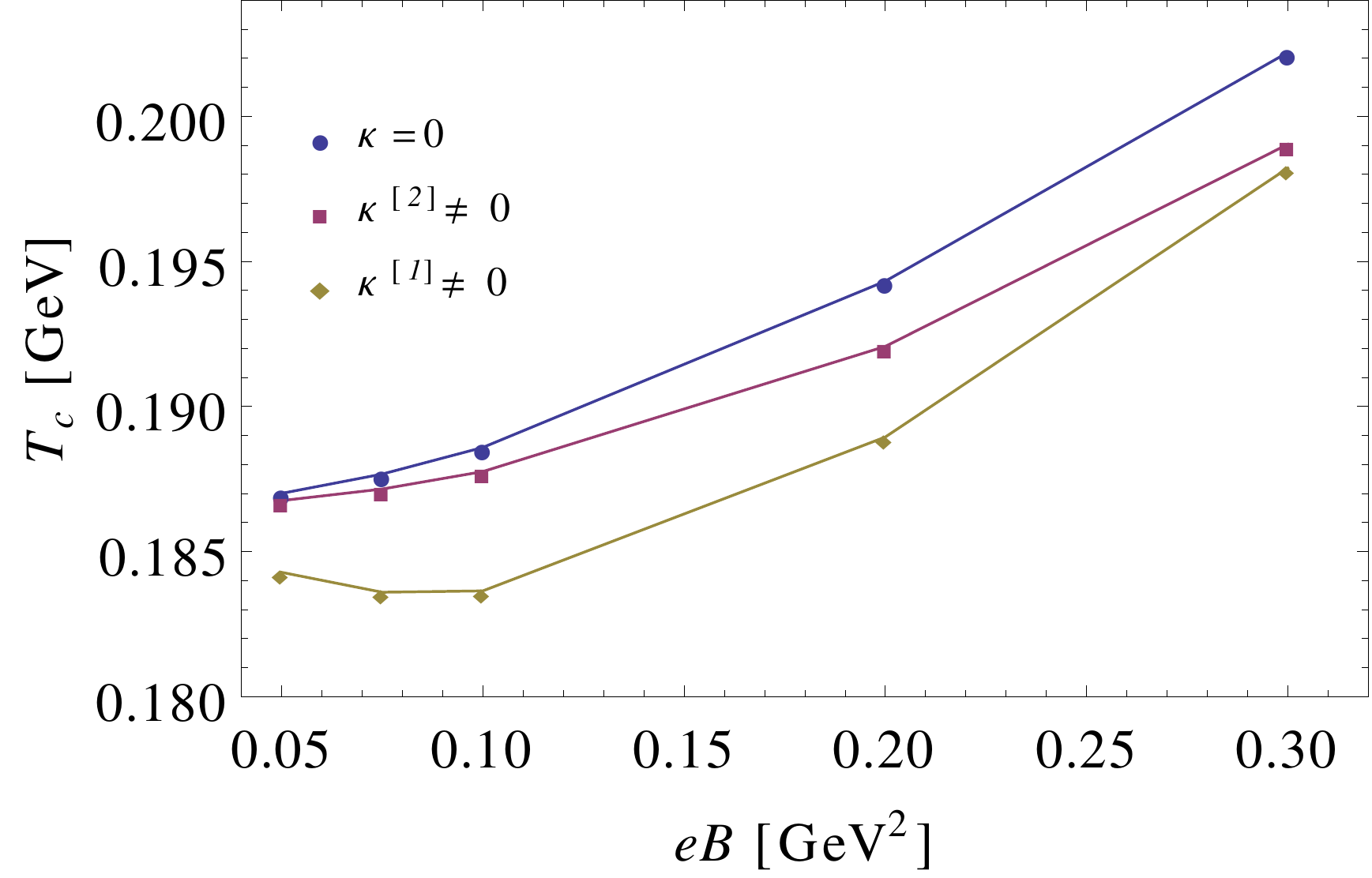}\\
\end{tabular}
\caption{The pseudocritical temperature for the chiral transition of magnetized quark matter as a function of the magnetic field for fixed values of AMM $\kappa$.}
\label{TcxB}
\end{figure}

To study how the pseudocritical temperature changes with the magnetic field when we consider fixed $\kappa_f$ values, we evaluate $-\partial M/\partial T$. 
The peak of each curve in Fig.~\ref{dmdt_kappa1} represents the transition point, $T_c$ with $\kappa^{[1]}$. We can see 
a smooth decreasing of $T_c$ at $eB\leqslant 0.1$ GeV$^2$. Beyond this, the pseudocritical
temperature grows as we increase the magnetic fields. On the other hand, when we fix $\kappa^{[2]}$, one observes the increase of the pseudocritical
temperature in Fig.\ref{dmdt_kappa2} for all values of magnetic fields. The behavior of the pseudocritical temperature, $T_c$, 
as function of the magnetic fields is clarified in the Fig.~\ref{TcxB}. We can clearly see that the values of $T_c$
are bigger in the $\kappa=0$ when 
compared to the and $\kappa^{[2]}$ and $\kappa^{[1]}$. 
 
We note that our results for $T_c$ as a function of the magnetic field do not present any kind of oscillation as some 
NJL predictions in the literature~\cite{Fayazbakhsh:2014mca,Chaudhuri:2019lbw}. 

\begin{figure}[h]
\begin{center}
\includegraphics[width=0.48\textwidth]{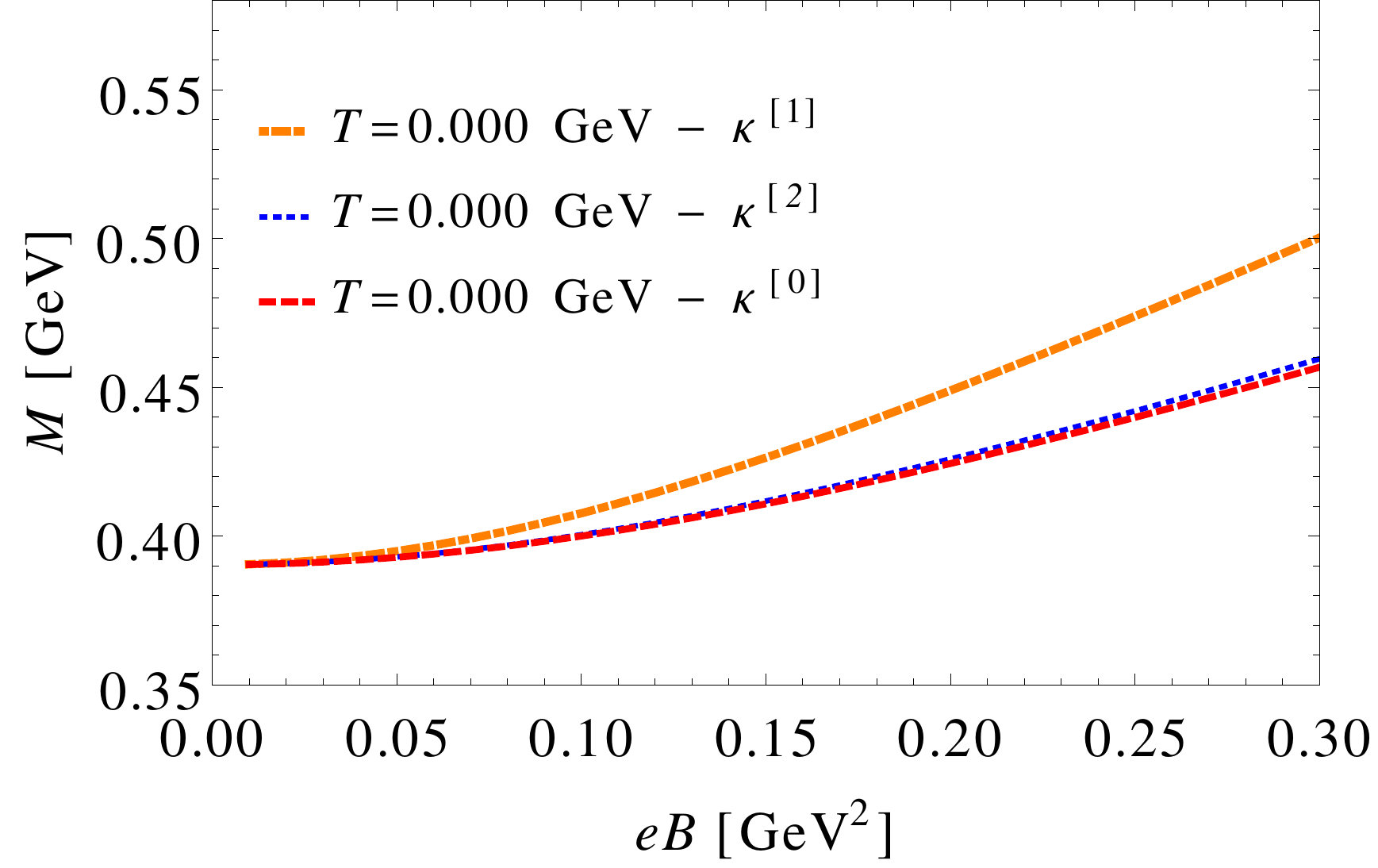}\vspace{0.5 cm}
\includegraphics[width=0.48\textwidth]{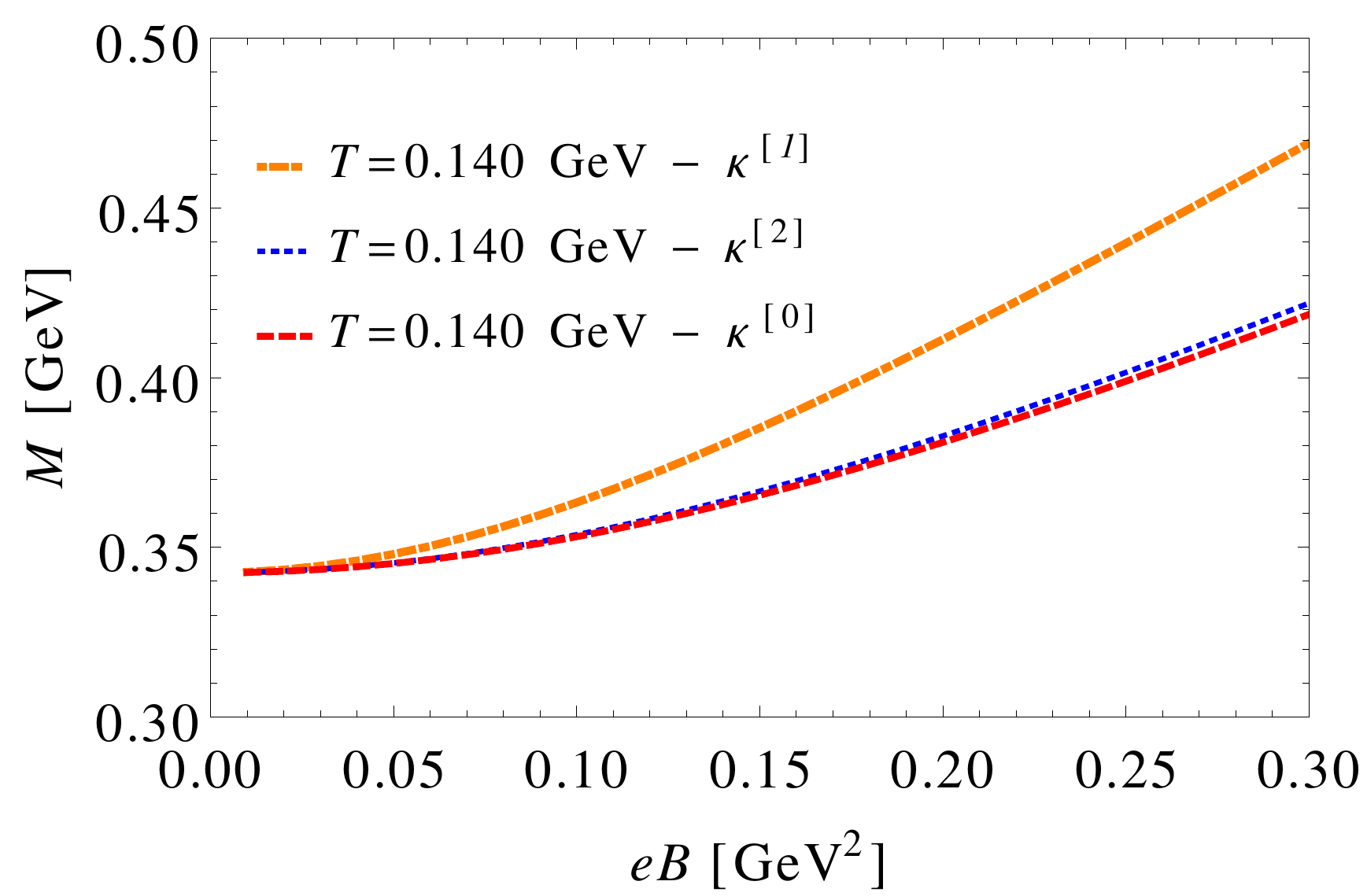}
\end{center}
\caption{Effective quark mass as function of the magnetic field for $T=0.05$ GeV (top panel) and $T=0.140$ GeV (low panel) with fixed values of $\kappa$.
}
\label{MxB_kappa}
\end{figure}

We also evaluate the effective quark mass as function of the magnetic fields with both sets of fixed $\kappa$ and $\kappa=0$, 
for the temperatures $T=0$ in the top panel of Fig.~\ref{MxB_kappa}
and $T=0.140$ GeV in the low panel. We can see that the values of the effective quark mass with $\kappa^{[1]}$ are almost the same as the one evaluated with
$\kappa^{[2]}$ and $\kappa=0$ for both temperatures considered in the weak magnetic field regime, e.g, $eB \lesssim 0.05$ GeV$^2$, and are always bigger than the ones with $\kappa^{[2]}$ and $\kappa=0$ in the strong magnetic field regime.
In our evaluations, there are no oscillations
in the effective quark masses, as observed in several evaluations in the $\kappa=0$ \cite{Avancini:2019wed} 
or with $\kappa\neq 0$ \cite{Fayazbakhsh:2014mca,Ghosh:2021dlo,Chaudhuri:2020lga,Chaudhuri:2019lbw}. 
These nonphysical oscillations observed in the literature for the constituent quark mass, pseudocritical temperature 
for chiral symmetry restoration and other physical quantities, can be traced to the fact that the regularization procedure 
depends explicitly on the magnetic field. We and other authors in the literature have shown in several works~\cite{Farias:2014eca,Farias:2016gmy,Avancini:2019wed,Menezes:2008qt,Avancini:2012ee,Menezes:2009uc,Avancini:2016fgq,Avancini:2017gck,Avancini:2018svs,Coppola:2017edn,Duarte:2015ppa,Allen:2015paa,Duarte:2016pgi,Avancini:2020xqe,Avancini:2015ady,Coppola:2018vkw,Tavares:2021fik,Aguirre:2021ljk,Aguirre:2020tiy} that these nonphysical oscillations disappear when the divergent terms are disentangled from the pure magnetic contributions by using the MFIR/VMR schemes.

\section{Conclusions}

In this work we have studied the effect of a constant anomalous magnetic moment in the SU(2) Nambu--Jona-Lasinio model. To this end,
we have employed the Schwinger {\it ansatz} representing the effect of AMM. Making use of VMR-type regularization, we obtain
a set of well defined expressions for the thermodynamical potential and the gap equation. For the effective quark mass, we
choose two sets of fixed values of AMM. For the set $\kappa^{[1]}$, which consider a sizable value of $\kappa_f$, 
we observe a smoothly decrease of $T_c$ for the magnetic field 
region, i.e., $eB \lesssim 0.1$ GeV$^2$. For the set $\kappa^{[2]}$, we observe the increase of $T_c$ for all magnetic field range adopted, whereas this
effect also take place with set $\kappa^{[1]}$ for the strong magnetic field region. Our results also shows
that the pseudocritical temperature is always bigger in the case of $\kappa=0$ than the $\kappa^{[1]}$ and $\kappa^{[2]}$ case. Therefore, one main effect of the AMM is to decrease the values of 
$T_c$ as we increase the value of $\kappa_f$.
The magnetic catalysis also holds for low temperatures with no oscillations, which is expected in the MFIR or VMR inspired regularization procedures.  
We hope that this work can clarify that these non-physical oscillations are an artifact of some 
regularization prescriptions that entangled the magnetic medium with the vacuum. Furthermore, these oscillations cannot be confused
with de Haas-van Alphen oscillations.
 
Our results when we consider nonvanishing quark AMM show that chiral symmetry restoration happens always as a smooth crossover and 
 never turns into a first order phase transition. We will report about the results exploring the AMM effects in the thermodynamical properties of magnetized, dense 
 and hot quark matter with VMR scheme in a future publication.

\appendix
\section{Vacuum magnetic regularization (VMR)}\label{appA}

To obtain the thermodynamical potential eq. (\ref{omegafull}) and the gap equation eq. (\ref{gap1}), we adopt a vacuum magnetic 
regularization scheme. First, the trigonometric part of the integrand of eq. (\ref{LAMM1}) should be expanded in Taylor series at $s\sim0$ as

\begin{eqnarray}
&& \coth(c_fB_f s)\sim 1+\frac{(c_fB_fs)^2}{2}+\mathcal{O}(s^4),\nonumber\\
&& \sinh(B_f s)^{-1}\sim \frac{1}{B_fs}-\frac{B_fs}{6}+\mathcal{O}(s^3),
\end{eqnarray}

\noindent where $c_f=1+\alpha_f$. Now we can see that, the integrand behaves like

\begin{eqnarray}
 \frac{\cosh(c_fB_fs)}{\sinh(B_f)}&\sim&\left[1+\frac{(c_fB_fs)^2}{2}\right]\left[\frac{1}{B_fs}-\frac{B_fs}{6}\right]\nonumber\\
                                          &=& \frac{1}{B_fs} +\frac{B_fs}{6}\left(1-3c_f^2\right)+\mathcal{O}(s^3),\label{div1}
\end{eqnarray}

\noindent  To eliminate the divergences, we add and subtract in the integrand the expansion eq. (\ref{div1}) in eq. (\ref{LAMM1})

\begin{eqnarray}
       \mathcal{L}&=&\sum_{f=u,d}\frac{N_c}{8\pi^2}\int_0^{\infty}\frac{ds}{s^3}e^{-s\mathcal{K}^2_{0f}}B_fs\left\{\frac{\cosh(c_fB_fs)}{\sinh(B_f)}-\frac{1}{B_fs} \right.\nonumber\\
                   &&\left.-\frac{(B_fs)^2}{6}\left(1-3c_f^2\right)\right\}+\Omega^{vac}+\Omega^{field},\nonumber\\
                   &=&\sum_{f=u,d}\frac{N_c}{8\pi^2}\int_0^{\infty}\frac{ds}{s^3}e^{-s\mathcal{K}^2_{0f}}\left\{\frac{B_fs\cosh(c_fB_fs)}{\sinh(B_f)}-1 \right.\nonumber\\
                   &&\left.-\frac{(B_fs)^2}{6}\left(1-3c_f^2\right)\right\}+\Omega^{vac}+\Omega^{field}\nonumber\\
                   &=&\Omega^{mag}+\Omega^{vac}+\Omega^{field}.\label{omegareg1}
\end{eqnarray}

\noindent The same procedure is adopted for the gap equation. The functions $\Omega^{vac}$ and $\Omega^{field}$ 
are regularized in the usual $3D$-cutoff scheme. The magnetic and thermo-magnetic contributions are finite.

\section{Constraints in the effective quark mass with AMM}\label{appB}

The application of a constant AMM value through the values of eq. (\ref{set1}) for $\kappa_f^{[1]}$ and eq. (\ref{set2})
for $\kappa_f^{[2]}$ limit the possible values of the effective quark mass. To see this, we
can work with the magnetic contribution, $h_f^{mag}$, of the gap equation, eq. (\ref{gap1}), as

\begin{eqnarray}
h^{mag}_f(M,eB)&=&\frac{MN_c}{4\pi^2}\int_{0}^{\infty}\frac{ds}{s^2}e^{-s\mathcal{K}_{0f}^2}\nonumber\\
               &&\times\left\{\frac{B_fs\cosh[c_fB_fs]}{\sinh(B_fs)}-1\right. \nonumber\\ 
               &&\left.-\frac{1}{6}\left[3(\alpha_f+1)-1\right](B_fs)^2\right\},\label{trig0}
\end{eqnarray}

\noindent where the pure trigonometric part of eq. (\ref{trig0}) non-regularized can be rewritten as

\begin{eqnarray}
 h^{mag}_f(M,eB)^{NR}&=&\frac{MN_c}{4\pi^2}\int_{0}^{\infty}\frac{ds}{s^2}e^{-s\mathcal{K}_{0f}^2}\nonumber\\
&&\times \left\{\frac{B_fs\cosh[c_fB_fs]}{\sinh(B_fs)}\right\}\nonumber
\end{eqnarray}

\noindent where we have adopted the superscript NR, that means "Non-Regularized". Making use of the following trigonometric relation

\begin{eqnarray}
 \cosh(a+b)=\cosh(a)\cosh(b)+\sinh(a)\sinh(b),
\end{eqnarray}

\noindent we obtain for $h^{mag}_f(M,eB)^{NR}$

\begin{eqnarray}
h^{mag}_f(M,eB)^{NR}&=&\frac{MN_c}{4\pi^2}\int_{0}^{\infty}\frac{ds}{s^2}e^{-s\mathcal{K}_{0f}^2}\nonumber\\
&&\times B_fs\left\{\frac{\cosh(\alpha_fB_fs)\cosh(B_fs)}{\sinh(B_fs)}\right.\nonumber\\
&&+\left.
\frac{\sinh(B_fs)\sinh(\alpha_fB_fs)}{\sinh(B_fs)}\right\}\nonumber\\ \label{trig2}
&=&\frac{MN_c}{4\pi^2}\int_{0}^{\infty}\frac{ds}{s^2}e^{-s\mathcal{K}_{0f}^2}B_fs\nonumber\\
&&\times \left\{\cosh(\alpha_fB_fs)\coth(B_fs)\right.\nonumber\\
&&+ \left.\sinh(\alpha_fB_fs)\right\}
\label{trig2}
\end{eqnarray}

Making use of the following relation 

\begin{eqnarray}
\sinh(x)=\frac{e^x-e^{-x}}{2},
\end{eqnarray}

\noindent we can rewrite the second integration, i.e., the integration of the $\sinh(\alpha_fB_fs)$, as

\begin{eqnarray}
&&\int_{0}^{\infty}\frac{ds}{s^2}e^{-s\mathcal{K}_{0f}^2}
\left\{ B_fs\sinh(\alpha_fB_fs)\right\}\nonumber \label{trig3}\nonumber\\
&&=\int_{0}^{\infty}\frac{ds}{s^2}e^{-s\mathcal{K}_{0f}^2}
\left\{ B_fs\frac{e^{\alpha_fB_fs}-e^{-\alpha_fB_fs}}{2}\right\}. \nonumber\\
\label{trig3}
\end{eqnarray}

It is clear from eq. (\ref{trig3}) that we will have a limit due to the $\alpha_f$ and $\kappa_f$ on the integration to avoid
a divergence. To see this, we will have the condition

\begin{eqnarray}
 -\mathcal{K}_{0,f}^2+\alpha_fB_f&<&0\nonumber\\
 -(M^2+(\kappa_fB_f)^2)+\alpha_fB_f&<&0\nonumber\\
  M^2&>&\alpha_fB_f-(\kappa_fB_f)^2
\end{eqnarray}

\noindent The above expression constrain the effective quark masses, $M$, for each of the flavors in gap equation.
As an estimative, if we take $\kappa^{[1]}_u$ in the above expression for $eB=0.1$ GeV$^2$, we obtain

\begin{eqnarray}
 M\gtrsim 125.693\hspace{0.4em}\text{MeV},
\end{eqnarray}

 The estimative for $\kappa^{[2]}_u$, we obtain
\begin{eqnarray}
 M \gtrsim 19.763\hspace{0.4em}\text{MeV}
\end{eqnarray}

\noindent both values are in agreement with the results in Fig.~\ref{MxT_kappa1} and Fig.~\ref{MxT_kappa2}.

\section*{Acknowledgments}

This work was partially supported by Conselho Nacional de Desenvolvimento Cient\'ifico 
e Tecno\-l\'o\-gico  (CNPq), Grants No. 309598/2020-6 (R.L.S.F.), No. 304518/2019-0 (S.S.A.); Coordena\c c\~{a}o  de 
Aperfei\c coamento de Pessoal de  N\'{\i}vel Superior - (CAPES) Finance  Code  001 (W.R.T); 
Funda\c{c}\~ao de Amparo \`a Pesquisa do Estado do Rio 
Grande do Sul (FAPERGS), Grants Nos. 19/2551- 0000690-0 and 19/2551-0001948-3 (R.L.S.F.); PET/MEC/Sisu (R.M.N); 
The work is also part of the project Instituto Nacional de Ci\^encia 
e Tecnologia - F\'isica Nuclear e Aplica\c{c}\~oes (INCT - FNA), Grant No. 464898/2014-5. 

\bibliographystyle{spphys}
\bibliography{AMM.bib}
\end{document}